\documentclass[12,preprint,superscriptaddress,nofootinbib,longbibliography]{revtex4-2}
\usepackage{xcolor}
\usepackage{graphicx}
\usepackage{dcolumn}
\usepackage{bm,amsfonts,amsthm,amsmath,amssymb,array}
\usepackage{bbm}
\usepackage[]{epsfig,graphics}
\usepackage{comment}
\usepackage[T1]{fontenc}
\usepackage{lipsum}
\usepackage[utf8]{inputenc}
\usepackage{ulem}
\usepackage{multirow}
\usepackage{color}
\usepackage{lastpage}
\usepackage[sort&compress]{natbib}
\usepackage{enumerate}
\usepackage{wasysym}
\usepackage{relsize}
\usepackage{physics}
\usepackage{mathrsfs}
\usepackage{tensor}
\usepackage{xfrac}
\usepackage{siunitx}
\usepackage{hyperref}
\usepackage{caption}
\usepackage{subcaption}

\hypersetup{
    unicode=false,          
    pdftoolbar=true,        
    pdfmenubar=true,        
    pdffitwindow=false,     
    pdfstartview={FitH},    
    pdftitle={My title},    
    pdfauthor={Author},     
    pdfsubject={Subject},   
    pdfcreator={Creator},   
    pdfproducer={Producer}, 
    pdfkeywords={keyword1} {key2} {key3}, 
    pdfnewwindow=true,      
    colorlinks=false,       
    linkcolor=red,          
    citecolor=green,        
    filecolor=magenta,      
    urlcolor=cyan           
}

\newenvironment{itemize*}
  {\begin{itemize}
    \setlength{\itemsep}{0pt}
    \setlength{\parskip}{0pt}}
  {\end{itemize}}

\newenvironment{enumerate*}
  {\begin{enumerate}
    \setlength{\itemsep}{0pt}
    \setlength{\parskip}{0pt}}
  {\end{enumerate}}

\newenvironment{description*}
  {\begin{description}
    \setlength{\itemsep}{0pt}
    \setlength{\parskip}{0pt}}
  {\end{description}}

\def\ben{\begin{enumerate*}}
\def\een{\end{enumerate*}}
\def\bi{\begin{itemize*}}
\def\ei{\end{itemize*}}
\def\bd{\begin{description*}}
\def\ed{\end{description*}}
\def\be{\begin{equation}}
\def\ee{\end{equation}}
\def\bea{\begin{eqnarray}}
\def\eea{\end{eqnarray}}
\def\bfl{\begin{flushleft}}
\def\efl{\end{flushleft}}
\def\bigskip{\;\;\;\;\;\;\;}

\def\ben{\begin{enumerate*}}
\def\een{\end{enumerate*}}
\def\bi{\begin{itemize*}}
\def\ei{\end{itemize*}}
\def\bd{\begin{description*}}
\def\ed{\end{description*}}
\def\be{\begin{equation}}
\def\ee{\end{equation}}
\def\bfl{\begin{flushleft}}
\def\efl{\end{flushleft}}

\newcommand{\Hcon}{\mathcal{H}}

\begin{document}

\title{Cosmological Perturbations from a New Approach to Inflation}

\author{Brandon Melcher} 
\email{brandon@appliedphysics.org}
\affiliation{Applied Physics, PBC 477 Madison Avenue, New York, NY 10022}

\author{Arnab Pradhan} \email{arpradha@syr.edu} 
\affiliation{Department of Physics, Syracuse
  University, Syracuse, NY 13244, USA}

\author{Scott Watson}
\email{gswatson@syr.edu}
\affiliation{Department of Physics, Syracuse
  University, Syracuse, NY 13244, USA}
  \affiliation{Department of Physics and Astronomy, University of South Carolina, Columbia, SC 29208, USA}

\date{\today}

\begin{abstract}
In a previous paper we proposed a new approach to the beginning of inflation -- a lingering universe \cite{Melcher:2023kpd}. The universe begins in a lingering state with a nearly vanishing Hubble parameter. This calls into question the absolute age of the universe, as the Hubble time can be nearly infinite. It also provides promise for addressing the initial singularity of inflation and issues with quantum field theory in de Sitter space-time. Such models arise in classical cosmologies with non-vanishing spatial curvature (inspired by PLANCK 2018 data), and independently by models that arise in string cosmology. In this paper, we consider the importance of cosmological perturbations for the stability of the lingering phase and how this influences cosmological observations. Our goal is to establish observables in this new paradigm for the origin of inflation which is in contrast to eternal inflation and cyclic cosmologies. We also address questions of stability and the transition to inflation.
\end{abstract}

\maketitle
\thispagestyle{empty}
\newpage

\section{Introduction: The Lingering Universe and Inflation}

In \cite{Melcher:2023kpd}, a new approach to the initial state of inflation was presented -- a lingering universe. These models are realized classically by considering the presence of non-trivial spatial curvature and a two-fluid system. One fluid corresponds to what becomes the inflationary sector, and the other has an equation of state corresponding to matter. As noted in \cite{Melcher:2023kpd} other possibilities can be realized, including a network of cosmic strings. In the same paper such a setup (without spatial curvature) was motivated within string theory with the key component being the string theory dilaton which leads to inflation.

In this paper, we address the issue of stability and the role of cosmological perturbations from the lingering phase\footnote{By lingering, we do not mean loitering in the sense of \cite{Brandenberger:2001kj}. However, these papers partially motivated our ideas}. Inflation is still the primary source of perturbations, but it is interesting that the lingering phase could lead (in principle) to observable transplanckian  effects. As a first step toward exploring this, we consider cosmological perturbations generated during the lingering phase. A key question is whether the lingering phase leads to observable predictions. In the classical case one possibility is a small, observable spatial curvature -- as suggested by the 2018 Planck data\footnote{We note that this is a speculative idea and criticisms of non-zero spatial curvature were discussed in \cite{Vagnozzi:2020dfn,Vagnozzi:2020rcz}} \cite{Planck:2018vyg}. 
In addition, the transplackian effects that may result from lingering where perturbations grow exponentially could be observationally interesting. 
In the string theory realization of lingering the Hagedorn phase can lead to predictions for the CMB -- or could place constraints on the Hagedorn phase as stated in \cite{Melcher:2023kpd}. 

Since CMB measurements provide the amplitude of inhomogeneities at a given scale \cite{Akrami:2018vks}, any growth of perturbations in the lingering phase must be compensated by a corresponding decay in the period following the lingering.  It is important to note that in the classical realization of our proposal, lingering demands the existence of positive curvature (i.e. the universe is closed), and so we will already have a lower bound on the duration of inflation.  The matching of inhomogeneities will provide a further restriction on the duration of lingering and inflation. In the string theory realization, there are still many issues to address -- in particular whether a Hagedorn transition occurs at higher energies, what sets the scales of the model (basically due to the dynamics of the dilaton resulting from its potential), and in general what is the initial time and period of the lingering.
Given these questions, it is still interesting that this would allow for calculations in quantum gravity that would avoid the difficulties of de Sitter space and an initial singularity problem. We note that whether there is an initial singularity problem for inflation has raised some disagreement in the literature. The authors of \cite{Kinney:2023urn} argue that the initial singularity of inflation does indeed exist arguing in favor of the BGV theorem resulting from geodesic incompleteness \cite{Borde:2001nh}, while the BGV result was challenged in \cite{Easson:2024fzn}. We emphasize that our motivation for this paper is not only focused on this issue of the singularity as discussed above. Our primary focus is a new approach for the beginning of inflation. 

Here we will focus on the classical model of \cite{Melcher:2023kpd}. 
Inflation is very effective at eliminating initial conditions and this is a challenge for establishing new observables and predictions of the initial state. Thus, we make modest first steps in addressing these issues, the stability of the solution, and what observables are possible (e.g. spatial curvature).

The rest of the paper is as follows.
In Section \ref{background} we explain the background evolution in the presence of positive spatial curvature in a two fluid classical system. 
In Section \ref{perts} we present our main results considering the evolution of the cosmological perturbations and the transition from lingering to an inflationary universe. We then conclude. 
\vspace{-0.2in}

\section{Background Evolution \label{background}}
In this section, we briefly summarize the background evolution with more details to be found in \cite{Melcher:2023kpd}.
To fully determine the background evolution, we work with conformal time ($a(\eta)d\eta=dt$) and suppose the presence of a non-zero spatial curvature, $K$. We propose that our spacetime possesses two fluids: a Standard-Model-like matter \cite{Sahni:1991ks}, ($w_s \geq 0$), and a fluid that scales slower than curvature: ($w_e\leq-1/3$) but does not violate the null energy condition \cite{Melcher:2023kpd}.  The equation of state of the fluids is given by ($p_i = w_i \rho_i$). The FLRW equations are then
\begin{align}
\mathcal{H}^2 &= \frac{\kappa^2}{3} a^2 \rho_{tot} - K \label{conFried1}\\
\mathcal{H}'+\frac{\mathcal{H}^2}{2} &= -\frac{1}{2}(\kappa^2 a^2 p_{tot} + K), \label{conFried2}
\end{align}
where $\mathcal{H}\equiv a'/a$, the prime is a derivative with respect to conformal time, and $\kappa^2\equiv8\pi G=m_p^{-2}$ where $m_p$ is the reduced Planck mass\footnote{We follow the notation and equations derived in \cite{Melcher:2023kpd}}.  We will find it more convenient to rescale our spacetime coordinates by the spatial curvature. We can do this without changing the form of the above equations since the FLRW metric is invariant under such a rescaling \cite{Melcher:2023kpd}.
The cosmology follows from solving \eqref{conFried1} and \eqref{conFried2} with $K=1$. To close this set of equations, we include the continuity equations for both fluids:
\be
\label{BCons}
\rho_i'+3\mathcal{H}(1+w_i)\rho_i = 0,
\ee
where $w_i = p_i/\rho_i$ is the equation of state for a given fluid. Given the assumptions on the equation of state of the fluids we have $\rho_i \sim a^{-3(1+w_i)}$. Using \eqref{conFried1} in \eqref{conFried2}, we obtain,
\be
\label{conFried2v2}
\mathcal{H}' = -\frac{\kappa^2 a^2}{6}\left(\rho_{tot} + 3 p_{tot}\right).
\ee
\vspace{-0.1in}
Another useful combination of \eqref{conFried1} and \eqref{conFried2} turns the Hubble equations into a second-order equation for $a$:
\be
\label{Scale2}
a'' + a = -\frac{\kappa^2 a^3}{6}(\rho_{tot} - 3 p_{tot}).
\ee

\begin{figure}[t!]
    \centering
         \includegraphics[width=0.7\textwidth]{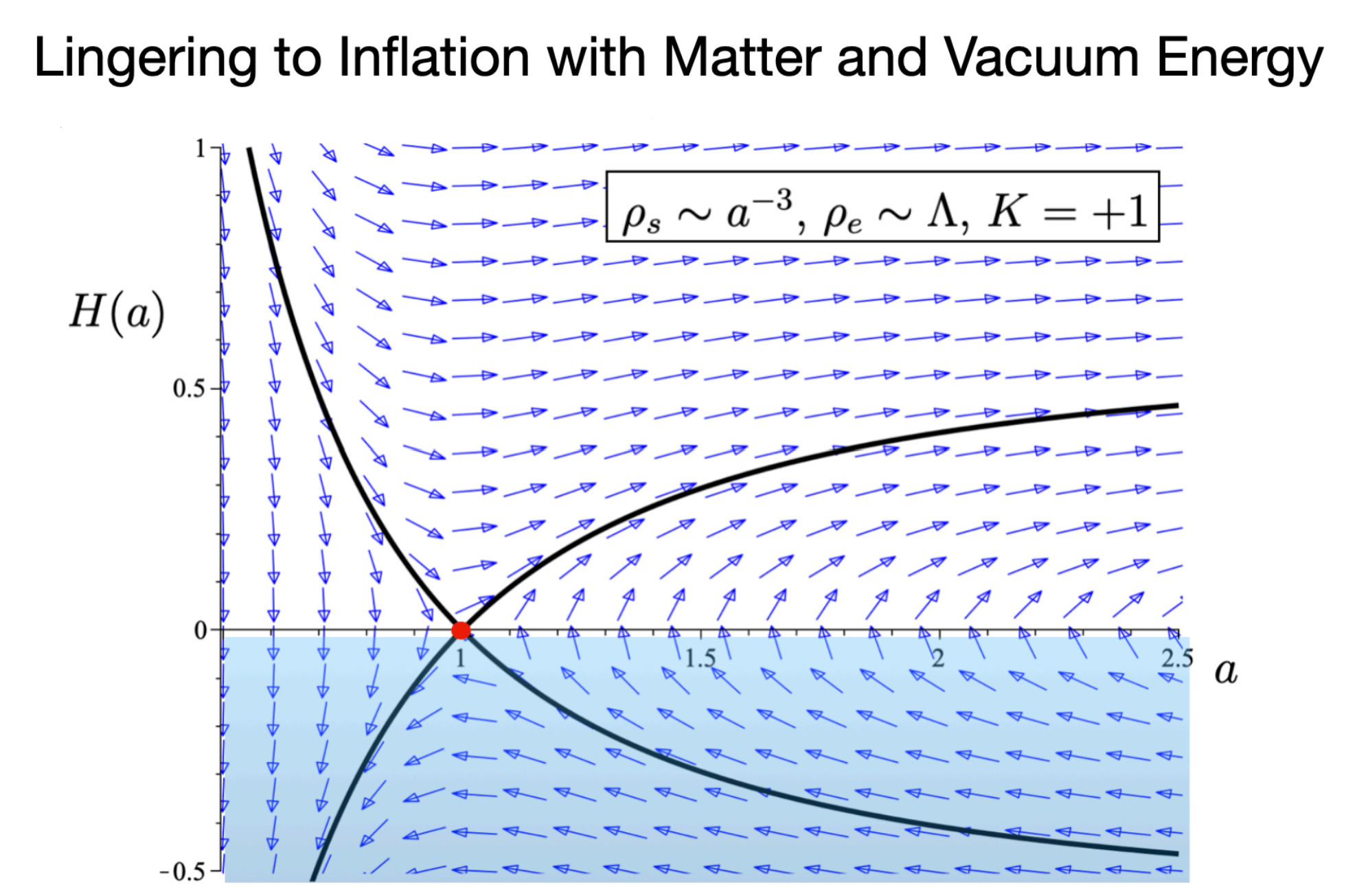}
    \caption{{ {\it Lingering Fixed Point.} The plot shows the flow lines in the Hubble parameter - scale factor phase space for a positively curved universe with standard matter and a fluid that scales like the cosmological constant. The red dot is an unstable (hyperbolic) fixed point in the evolution, it corresponds to initial conditions for exactly reaching the lingering phase. The black curves in both figures correspond to the enforcement of the Hubble constraint (energy conservation). The blue region emphasizes we are interested in a positive expansion rate.} \label{fig:flow}}
\end{figure}
\subsection{Lingering} To obtain an exact lingering phase we require 
\begin{align}
\frac{\overline{\rho}_s}{a^m} + \frac{\overline{\rho}_e}{a^n} - \frac{3}{\kappa^2 a^2} = 0, \label{adot0} \\
(m - 2) \frac{\overline{\rho}_s}{a^m} + (n - 2) \frac{\overline{\rho}_e}{a^n} = 0, \label{addot0}
\end{align}
where we have defined $3(1 +  w_e) = n$ and $3(1 +  w_s) = m$, which imply that $0\leq n \leq 2$ and $m \geq 3$. One should note that we have defined $\rho_i = \overline{\rho}_i a^{-3(1+w_i)}$. We have from \eqref{adot0} and \eqref{addot0} that $a'=a''=0$. We define a perfectly lingering spacetime as one where the first and second time derivatives of the scale factor vanish \textit{for all time}. 

This suggests that rather than finding specific values of $\overline{\rho}_s$ and $a$ such that the universe stalls, the important question is actually what values of $\overline{\rho}_s$ and $\overline{\rho}_e$ are allowed by a given scale factor and curvature. Therefore, we find that for

\be
\label{LoitEnDens}
\overline{\rho}_s = \overline{\rho}_s^\ast \equiv \frac{2-n}{m-n} \frac{3}{\kappa^2} a_\ast^{m-2} \quad \mbox{and}, \quad \overline{\rho}_e = \overline{\rho}_e^\ast \equiv \frac{m-2}{m-n} \frac{3}{\kappa^2} a_\ast^{n-2},
\ee
the universe will linger at scale factor $a_\ast$ as seen in Fig. \ref{fig:flow}. We also remind the reader that it was from the phase space analysis in \cite{Melcher:2023kpd} that the lingering point is a hyperbolic fixed point that leads asymptotically to inflation. Also, notice that our assumptions on the equations of state imply that both energy densities remain positive. 
We note that the $n=2$ case is particularly special.  In that case, the universe can linger at \textit{any} scale factor as long as $\overline{\rho}_s = 0$ and $\overline{\rho}_e = 3 \kappa^{-2}$.

Our interest lies in the amount of matter for which the lingering phase ends in a finite amount of time, i.e. $\overline{\rho}_s \neq \overline{\rho}_s^\ast$ and $\overline{\rho}_e \neq \overline{\rho}_e^\ast$. We characterize the length of the lingering phase by tracking the ``nearby'' scale factor trajectories\footnote{Again, we emphasize this is a phenomenological approach, the ultimate goal would be to calculate the exact numbers in a fundamental theory.}.  To this end, we suppose that $\overline{\rho}_e = \overline{\rho}_e^\ast(1 + \Delta_e)$, and $a(\eta) = a_\ast (1 + \Delta(\eta))$.  If the exotic matter scales like curvature the lingering amount of clustering matter is zero.  In that case, we can't parameterize the deviation of the energy density as a fraction of what is required for lingering.  There are two possible changes to the clustering matter: $\overline{\rho}_s = \overline{\rho}_s^\ast(1 + \Delta_s) + \Delta\rho_s$, where $\Delta_s$ will vanish in the $n=2$ case, and we can just set $\Delta\rho_s=0$ otherwise. Expanding \eqref{conFried1} and \eqref{conFried2} to first order in $\Delta$, $\Delta_e$, and $\Delta_s$ or $\Delta\rho_s$. For $n\neq2$,
\begin{align}
(m - 2) \Delta_e + (2 - n) \Delta_s =& 0, \label{Hub1O1} \\
\Delta'' + \frac{1}{2} \bigg( (m - 2) (n - 2) \Delta + \frac{(m - 2) (n - 3)}{m - n} \Delta_e - \frac{(m - 3) (n - 2)}{m - n} \Delta_s \bigg) =& 0. \label{Hub2O1}
\end{align}
The first of these equations ensures that $\kappa^2 a_\ast^2 \rho_{tot}/3 =  K = 1$. Assuming the new trajectory exactly matches the perfect lingering solution at some starting point $\eta_0$, the above equations are solved by
\be
\label{LoitScaleGen}
\Delta = \frac{\Delta_e}{2 - n} \bigg[ \cosh\bigg(\sqrt{\frac{1}{2}(m-2)(2-n)} \Delta\eta\bigg) - 1 \bigg],
\ee
where $\Delta\eta = \eta - \eta_0$. When $n=2$, we have to deal with a perturbation in the Standard Model fluid as an ``absolute'' energy density. The above equations become
\begin{align}
\Delta_e + \frac{\kappa^2}{3 a_\ast^{m - 2}} \Delta\rho_s =& 0, \label{Hub1N2O1} \\
\Delta'' + \frac{1}{6}\bigg( \frac{\kappa^2}{a_\ast^{m - 2}} (m - 3) \Delta\rho_s - 3 \Delta_e \bigg) =& 0, \label{Hub2N2O1}
\end{align}
which implies
\be
\label{LoitScale2}
\Delta = \frac{m - 2}{4}\Delta_e \Delta\eta^2.
\ee
One can pass from the $n\neq2$ solution to the $n=2$ solution via the formal limit $n\rightarrow 2$.

This linear approximation breaks down when $\Delta$ approaches one; this signals the exit from the lingering phase.  Via inspection of the solutions above, one finds that with energy densities closer to their exact lingering values, the longer the duration of the quasi-lingering phase lasts, as one should expect. We can estimate the end of lingering as
\be
\Delta \eta_{e} \simeq \sqrt{\frac{2}{(m - 2) (2 - n)}} \ln \bigg((2 - n) \frac{2}{\Delta_e} \bigg)
\ee
when $n < 2$. We assumed that $\Delta_e \ll 1$ to arrive at the above expression. When $n = 2$, 
\be
\Delta \eta_{e} = \frac{2}{\sqrt{(m - 2) \Delta_e}}. 
\ee
Both of these time durations obey $\Delta\eta_e\rightarrow\infty$ for $\Delta_e\rightarrow0$, which is the perfectly lingering limit.

To this point, we made no assumptions on $\Delta_e$, $\Delta_s$, or $\Delta\rho_s$. The forms of \eqref{Hub1O1} and \eqref{Hub1N2O1}, as well as the ranges for $m$ and $n$, other than they respect these values respect the null energy condition.  The universe, desiring to stall at the same scale factor as in its perfect lingering form, must compensate any additional energy in a given fluid by a corresponding loss in the other. When we have reduced the amount of energy in the exotic fluid, the universe tends to recollapse, as expected for a closed universe filled with ``standard'' matter.

The $n=2$ case is again special -- as an exact lingering scenario requires zero standard matter to achieve a steady-state universe. Thus, any additional exotic matter inserted to break lingering requires the corresponding reduction of standard matter. 
The universe needs negative energy density in the standard matter sector to obtain an initially lingering solution followed by an exponentially expanding inflation-like solution. 

\subsection{Post-lingering} By assumption, the matter fluid dilutes with the cosmic expansion the fastest.  Neglecting the matter contribution then provides us with the equations that determine the evolution of the scale factor in the post-lingering era.  In this limit, \eqref{Scale2} takes the form
\be
\label{PLDifEq1}
a'' + a = \frac{(m-2)(n-4)}{(n-m)} \frac{(1+\Delta_e)}{2 a_\ast^{2-n}} a^{3-n} \equiv C a^{3-n},
\ee
where we have inserted the ans\"atz for $\overline{\rho}_e$ discussed above.  The Hubble equation then takes the form 
\be
\label{PLDifEq2}
\Hcon^2 = \frac{2(n-m)C}{(3-n)(m-2)(n-4)} a^{2-n} - 1.
\ee
The equations can be solved for $a(\eta)$ for the relevant values of $n$. We denote by $\Delta \eta_{pl}$ the conformal time duration since the post-lingering transition. The general solution for \eqref{PLDifEq1} and \eqref{PLDifEq2} is
\be
\label{PostLoitScaleGen}
a = \Bigg( -\frac{m-2}{n-m} \frac{1+\Delta_e}{a_\ast^{2-n}}\Bigg)^{1/(n-2)} \sin(\frac{n-2}{2}\Delta\eta_{pl} + \eta_0)^{2/(n-2)},
\ee
where $\eta_0$ is the remaining constant of integration set by the initial value of the scale factor at the start of the post-lingering phase of evolution. From this solution, we need to address the $n=2$ case separately. For $n=2$ one has
\be
\label{Scalen2}
a_2(\eta) = a_{pl} \exp\left(\Delta \eta_{pl} \sqrt{\Delta_e}\right),
\ee
where \eqref{PLDifEq2} implies we must choose either the exponentially expanding or contracting solution of \eqref{PLDifEq1}.
Given the background cosmology we now turn to the issue of cosmological perturbations. 

\section{Perturbation Evolution \label{perts}}
\subsection{Cosmological Perturbations}
For hydrodynamical perturbations in this FLRW universe with curvature we find that in Newtonian gauge the equations are\footnote{We follow the notation of \cite{Mukhanov:1990me}}
\begin{align}
3\Hcon \Phi'+3\left(\Hcon^2-1\right)\Phi - \nabla^2 \Phi &= -\frac{\kappa^2}{2}a^2\sum_i \delta\rho_i, \\
-\nabla^2 \left(\Phi'+\Hcon \Phi\right) &= \frac{\kappa^2}{2}a^2\sum_i (\rho_i+p_i)\theta_i,  \\
\Phi''+3\Hcon\Phi'+\left(2\Hcon'+\Hcon^2-1\right)\Phi &= \frac{\kappa^2}{2}a^2\sum_i \delta p_i, \\
\delta \rho_i' - 2\rho_i'\Phi + 3\Hcon(\delta\rho_i+\delta p_i) + (\rho_i+p_i)(\theta_i - 6\Hcon \Phi - 3 \Phi') &= 0 \\
\left[(\rho_i + p_i) \theta_i\right]' + 4\Hcon(\rho_i + p_i) \theta_i + \nabla^2 \delta p_i + (\rho_i + p_i) \nabla^2 \Phi &= 0.
\end{align}
In the above, we have assumed no anisotropic stress so that we can characterize the scalar metric perturbations solely in terms of $\Phi$.  We have defined $\delta\rho_i$ and $\delta p_i$ to be the deviation from the background energy density and pressures, respectively.  $\theta_i$ is three-dimensional Laplacian of the scalar velocity perturbation potential. 

\subsection{Eigenmode decomposition}
Typical analysis of cosmological perturbations proceeds by expanding the perturbation functions in Fourier modes due to $e^{i p\cdot x}$ being a complete set of eigenfunctions for the {\it flat} space Laplacian.  As required for lingering, our universe possesses non-trivial spatial curvature. Following \cite{Mukhanov:1990me,Lifshitz:1963ps,Birrell:1982ix}, one finds a useful basis is
\be
X(\vec{x},\eta) = \sum_\mathbf{k} X_\mathbf{k} (\eta) L_\mathbf{k}(\vec{x}),
\ee
where $L_\mathbf{k}$ form a complete set of eigenfunctions of the Laplacian. While in flat space, the formal sum above can be taken over a continuous variable, thereby transforming it to a Fourier transform. If the spatial curvature is positive, the sum remains discrete. The eigenfunctions are
\be
L_\mathbf{k}(\vec{x}) \propto \Pi_{k,J}(\chi) Y^M_J(\theta,\phi),
\ee
where $Y^M_J$ are the standard spherical harmonics, and $\Pi_{k,J}$ can be represented by \cite{Birrell:1982ix,Lifshitz:1963ps}
\be
\label{EigenFunctionClosed}
\Pi_{k,J} = (\sin \chi)^J \Big(\frac{\dd}{\dd \cos\chi}\Big)^{J+1} \cos k \chi.
\ee
Note that $\chi$ can be thought of as a radial coordinate for the eigenfunctions.  Since $\chi \in (0,2\pi)$, we confirm that $k$ must be discrete (given that it is a compact space), as are $J$ and $M$. Thus $\mathbf{k} = (k, J, M)$, where $k \in \mathbb{N}$, $J \in (0, k-1)$, and $M \in (-J, J)$. Inspection of the above eigenfunctions allows one to identify $k$ as the curved space equivalent of wave-number. Integration over three-space sets the normalization for a given eigenfunction $L_\mathbf{k}$, but for our current investigations, the eigenvalues of these functions are the only relevant quantity. One finds
\be
\nabla^2 L_\mathbf{k} = - (k^2 - 1) L_\mathbf{k}.
\ee
To operate in the momentum space mode expansion in a universe with positive spatial curvature, we replace $\nabla^2 X \rightarrow -(k^2 - 1) X_\mathbf{k}$ for all perturbative fields $X$. Furthermore, we define the adiabatic sound speed of the perturbations as $\delta p_i = c_{s,i}^2 \delta \rho_i$ and use density contrasts, defined as $\delta_i \equiv \delta \rho_i/\rho_i$, to write the perturbation equations as:
\begin{align}
3\Hcon \Phi' +  \left(3 \Hcon^2 + k^2 - 4\right) \Phi &= -\frac{\kappa^2 a^2}{2} \left(\rho_s \delta_s + \rho_e \delta_e \right) \label{phitt} \\
(k^2 - 1) \left(\Phi' + \Hcon \Phi\right) &= \frac{\kappa^2 a^2}{2} \left( (1 + w_s)\rho_s \theta_s + (1+\gamma) \rho_e \theta_e\right) \label{phitx} \\
\Phi'' + 3 \Hcon \Phi' + \left(2\Hcon' + \Hcon^2 - 1\right) \Phi & = \frac{\kappa^2 a^2}{2} \left(c_{s,s}^2 \rho_s \delta_s + c_{s,e}^2 \rho_e \delta_e\right) \label{phixx} \\
\delta_i' + 3\Hcon(c_{s,i}^2 - w_i)\delta_i + (1+w_i) (\theta_i - 3 \Phi') &=0 \label{ECons} \\
\theta_i' +(1 - 3 w_i) \Hcon \theta_i - (k^2- 1) \left(\frac{c_{s,i}^2}{1 + w_i} \delta_i + \Phi\right) &= 0. \label{ProbPos}
\end{align}
As long as $w_i' = 0$ for each fluid, we can simplify the above a bit further by setting $c_{s,i}^2 = w_i$, which we assume for the remainder of the paper. Eq's \ref{ECons} and \ref{ProbPos} can be combined to give
\be
\delta^{''}_i +(1-3w_i)\Hcon\delta^{'}_i+(k^2-1)c_{s,i}^2\delta_i+(1+w_i)(k^2-1)\Phi-3(1+w_i)(1-3w_i)\Hcon\Phi^{'}-3(1+w_i)\Phi^{''}=0
\label{denconsttime}
\ee
which we use to study the time evolution of the density contrasts. One can obtain a similar equation for $\theta$.

\subsection{Exact Lingering Perturbations}
When $\Hcon$ and $\Hcon'$ are exactly zero, we obtain the perturbations for a universe at the exact static point discussed above. Thus, only the perturbative quantities are dynamic. To simplify notation we define,
\be
S \equiv \kappa^2 a^2 \rho_s, \quad E \equiv \kappa^2 a^2 \rho_e.
\ee
The background equations then become
\be
S + E = 3, \quad w_s S + w_e E = -1.
\ee
The perturbations are described by
\begin{align}
(k^2 - 4) \Phi &= -\frac{1}{2} ( S \delta_s + E \delta_e ), \label{ExLi1} \\
(k^2 - 1) \Phi' &= \frac{1}{2} ( (1+w_s) S \theta_s + (1+w_e) E \theta_e), \label{ExLi2} \\
\Phi'' - \Phi &= \frac{1}{2} (w_s S \delta_s + w_e E \delta_e), \label{ExLi3} \\
{\delta_i}' + (1 + w_i) (\theta_i - 3 \Phi') &= 0, \label{ExLi45} \\
{\theta_i}' - (k^2 - 1)(\frac{w_i}{1 + w_i} \delta_i + \Phi) &= 0, \label{ExLi67}
\end{align}
where $i$ is either standard matter ($s$) or `exotic' ($e$) depends on which fluid's perturbations are considered. \\ \indent 
Our perturbations possess a discrete spectrum due to the closed universe assumption. From the above equations, the modes with $k = 1,2$ clearly require extra attention. The first of these special modes are zero modes, i.e. they re-parameterize the background. Due to the background being an unstable fixed point in phase space, we expect that  $k = 1$ modes will have exponential growth for all equations of state. The $k=2$ mode vanishes in the linear perturbation theory by the background assumption of isotropy (the dipole vanishes given the linearity of the perturbations). However, this does raise an important issue with our approach and one that has been a criticism of inflation. That is, it would seem one must assume a smooth (homogeneous and isotropic) patch for inflation to occur in the first place. Here we make this assumption, and leave this consideration to future work, particularity in the string theory realization of a lingering initial phase.  \\ \indent
The simplest case we can consider is that of the exotic fluid being a cosmological constant\footnote{Providing no perturbations given time-translation invariance.}, which implies that $\delta_e = \theta_e = 0$ for all time. While the other cases we want to consider will involve 
evolving fluids, yet the equations remain the same throughout the exact lingering analysis given our assumptions. We discuss our calculation in-detail for the cosmological constant case to prepare the reader for the more complicated versions of the same assessment. 

For this special case we have
\begin{align}
(k^2 - 4) \Phi &= -\frac{1}{2} S \delta_s , \label{ExLiCC1} \\
(k^2 - 1) \Phi' &= \frac{1}{2} (1+w_s) S \theta_s , \label{ExLiCC2} \\
\Phi'' - \Phi &= \frac{1}{2} w_s S \delta_s , \label{ExLiCC3} \\
{\delta_s}' + (1 + w_s) (\theta_s - 3 \Phi') &= 0, \label{ExLiCC4} \\
{\theta_s}' - (k^2 - 1)(\frac{w_s}{1 + w_s} \delta_s + \Phi) &= 0. \label{ExLiCC5}
\end{align}

For the $k = 1$ mode, it follows immediately that the velocity divergence is constant and zero ($\theta_s = 0$). The background tells us that $1 + w_s = 2/s$, and so \eqref{ExLiCC1} and \eqref{ExLiCC4} are actually equivalent. Using \eqref{ExLiCC1} in \eqref{ExLiCC3}, we find that 
\be
\Phi'' - (1 + 3 w_s) \Phi = 0,
\ee
which leads to real exponential solutions for $w_s > -1/3$, as expected.

Moving onto $k = 2$, we find that $\delta_s = 0$. 
This again vanishes due to the assumptions of the background and that we are using linear perturbations. 
For $k \geq 3$, we first take the conformal-time derivative of \eqref{ExLiCC4} and use \eqref{ExLiCC5} to eliminate $\theta_s$. Then we can combine \eqref{ExLiCC1} and \eqref{ExLiCC3} to eliminate $\Phi$. The resulting dynamical equation for ``standard'' matter perturbations is\footnote{Alternatively, one could perform similar replacements to find a second-order equation for $\Phi$ and the same stability condition analysis would hold.}
\be
{\delta_s}'' + ((k^2 + 2) w_s - 1) \delta_s = 0.
\ee
This mode will only oscillate in time if 
\be
w_s > \frac{1}{k^2 + 2}.
\ee
Since the right side of this inequality is a decreasing function of $k$, we find that if $w_s > 1/11$, all higher-order modes will be purely oscillatory\footnote{We note that 
this result agrees with the results in \cite{Sahni:1991ks} for a late-time stalled universe if we take the limit $w_s =0$ where lingering was considered in a different context.}.

The above analysis suggests that while a purely static universe containing a cosmological constant and dust would inevitably develop clusters of matter that ruins the lingering phase, a combination of these and radiation fluids would be able to exist for an indeterminate period of time. This assumes that the energy densities only evolve through cosmic expansion.

Moving on to the case where $w_e = -2/3$, we again look at the first two modes of the perturbative expansion in more detail. When $k = 1$, we have a constant velocity divergence for both fluids, similarly to the cosmological constant case. We can eliminate the $\Phi$-dependence in the $\delta$ equations similar to above. It is more convenient to solve for the quantities
\begin{align}
\Delta \equiv \delta_s - \delta_e, \quad \mbox{and} \quad T \equiv \delta_s + \delta_e.
\end{align}
Their dynamics are dictated by
\begin{align}
\Delta'' &= \frac{1}{2}(1 + 3 w_s) \Delta, \\
T'' &= \frac{1 + 3 w_s}{2} \frac{4 + 3 w_s}{2 + 3 w_s} \Delta
\end{align}
Since $w_s \geq 0$, the fluid perturbations clearly diverge, which in turn causes the metric perturbation to explode. Again, this is the expected instability of the background.

At the $k=2$ level, we now find that the density perturbations are not algebraically independent. 
Equation \eqref{ExLi1} implies that
\be
\delta_e = - \frac{\delta_s}{1 + 3 w_s}.
\ee
Using this result in conjunction with the same replacement process as above, we find that the fluid perturbations obey
\be
{\delta_s}'' + \frac{3}{2}(w_s - 1) \delta_s = 0,
\ee
which will blow-up for $w_s < 1$. This is again suggestive of the importance of our fluid sectors respecting the Null Energy Condition. Otherwise, we would encounter critical instabilities. Fortunately, when a model like this is realized in a fundamental theory one finds this property is typically respected on global scales \cite{Adams:2006sv,Kaloper:2007pw,Melcher:2023kpd}. \\ 
For the higher-order modes, a somewhat tedious calculation allows one to show that the fluid perturbations are governed by 
\be
{\delta_s}'' = A \delta_s + B \delta_e, \quad 
{\delta_e}'' = C \delta_e + D \delta_s,
\ee
where
\begin{align}
A &= \frac{3 - 4 (k^2 - 4) w_s + (15 - k^2) w_s^2}{4 + 6 w_s}, \\
B &= -3 \frac{1 + 4 w_s + 3 w_s^2}{4 + 6 w_s}, \\
C &= \frac{4 k^2 (2 + 3 w_s) - 21 w_s - 11}{6(2 + 3 w_s)}, \\
D &= \frac{1 + 3 w_s}{2(2 + 3 w_s)}.
\end{align}
These equations are solved by $e^{\pm \sqrt{\omega_\pm^2} \eta}$, where
\be
\omega_\pm^2 = A + C \pm \sqrt{(A - C)^2 + 4 B D}.
\ee  
\subsection{Total stress-energy perturbation variables}
The previous equations can in principle be solved for all the perturbative quantities. To calculate the power spectrum of the curvature perturbation  we indeed require all of that information, but solving all equations simultaneously is challenging -- especially since the Hubble parameter does not completely vanish during the lingering phase. 
Thus, we first collect the fluid quantities and access the evolution of the metric perturbation before knowing exactly how the separate fluids evolve themselves since the right-hand-sides of the Einstein equations can be simplified in terms of total stress-energy perturbation variables.  These are defined as 
\begin{align}
\delta_T &= \Omega_s \delta_s + \Omega_e \delta_e \\
\theta_T &=\frac{1+w_s}{1+w_{eff}} \Omega_s \theta_s + \frac{1+\gamma}{1+w_{eff}} \Omega_e \theta_e \\
w_{eff} &= w_s \Omega_s + \gamma \Omega_e \\
c_{s,T}^2 &= \frac{1}{\delta_T}\left(c_{s,s}^2 \Omega_s \delta_s + c_{s,e}^2 \Omega_e \delta_e\right),
\end{align}
where we defined $\Omega_c = (3 \Hcon^2)^{-1} (\kappa^2 a^2 \rho_c)$, $\Omega_e = (3 \Hcon^2)^{-1} (\kappa^2 a^2 \rho_e)$, and $\Omega_K = (\Hcon^2)^{-1} $, which implies that $\Omega_c + \Omega_e - \Omega_K = 1$.  This allows us to write the Einstein equations as
\begin{align}
3\Hcon \Phi' + \left(3\Hcon^2 + k^2 - 4\right) \Phi &= -\frac{3}{2} \Hcon^2 \delta_T \label{phitt2} \\
(k^2 - 1) \left(\Phi' + \Hcon \Phi\right) &= \frac{3}{2}\Hcon^2 (1 + w_{eff})\theta_T \label{phitx2} \\
\Phi'' + 3 \Hcon \Phi' + \left(2\Hcon' + \Hcon^2 - 1\right) \Phi & = \frac{3}{2}\Hcon^2 c_{s,T}^2 \delta_T .\label{phixx2} 
\end{align}
We can combine the two Einstein equations with $\delta_T$ above to find that
\be
\label{GenEinCombo}
\Phi'' + 3 \Hcon (1 + c_{s,T}^2) \Phi' + \left(3 \Hcon^2 (c_{s,T}^2 - w_{eff}) + c_{s,T}^2 k^2 - 2 (2c_{s,T}^2 - 1)\right) \Phi = 0, 
\ee
where, with our definition of $w_{eff}$ and \eqref{conFried2v2}, we used
\be
\Hcon' = -\frac{1}{2} \left((1 + 3 w_{eff}) \Hcon^2 + 1\right).
\ee
By defining a new variable
\be
\label{phiconv}
u_n = \frac{a}{\kappa} \sqrt{\frac{2}{\mathcal{D}_n}} \Phi_n,
\ee
where $n$ characterizes the exotic fluid as described previously, we can simplify the above to
\be
\label{ptbeq}
u_n'' + \left(c_{s,T}^2 (k^2 - 1) - \frac{z_n''}{z_n}\right)u_n = 0,
\ee
where we have defined 
\be
\label{ScalarTDepMass}
z_n \equiv \frac{\Hcon}{a} \sqrt{\frac{3}{2 \mathcal{D}_n}}, \quad \mbox{and} \quad \mathcal{D}_n \equiv \Hcon_n^2 - \Hcon_n' + 1.
\ee
Neither $\Phi$ nor $\delta_T$ lead to observational signatures. This role is played by $\zeta$:
\be
\zeta = -\Phi + \frac{\delta \rho_{tot}}{3(\rho_{tot}+p_{tot})}.
\ee
Using the collective definitions above, along with the Friedmann equations we find
\be
\zeta = -\Phi + \frac{\Hcon^2+1}{2(\Hcon^2 - \Hcon' + 1)} \delta_T = -\Phi - \frac{(\Hcon^2+1)(3\Hcon \Phi' + \left( 3\Hcon^2 + k^2 - 4 \right) \Phi)}{\Hcon^2(\Hcon^2 - \Hcon' + 1)},
\ee
where we used Eq. \ref{phitt2} for the last equality.


The above equation therefore provides the relevant regimes of evolution: for $z''/z \gg c_{s,T}^2 (k^2-1)$ there is an effective tachyonic mass and
\be
u = C_1 z + C_2 z \int \frac{\dd \eta'}{z^2},
\ee
where $C_1$ and $C_2$ are integration constants. In the opposite limit,
\be
u = C_1 e^{i \sqrt{c_{s,T}^2 (k^2 - 1)}\eta} + C_2 e^{-i \sqrt{c_{s,T}^2 (k^2 - 1)} \eta },
\ee
where we have implicitly assumed that the total sound speed of the fluid varies slowly in the background regimes of interest. The background evolution of cosmic expansion thus plays the role of a time-dependent frequency in the equation for the perturbative quantity $u_n$.

Since $k^2-1>0$, the sign of $c_{s,T}^2$ controls the behavior of $u_n$. With assumptions stated above, we find
\be
c_{s,T}^2 = \frac{w_s \Omega_s \delta_s + w_e \Omega_e \delta_e}{\Omega_s \delta_s + \Omega_e \delta_e} = \frac{w_s Y + w_e}{Y + 1},
\ee
where we defined
\be
Y \equiv \frac{\Omega_s \delta_s}{\Omega_e \delta_e} = \frac{\rho_s \delta_s}{\rho_e \delta_e} = a^{n-m} \frac{\overline{\rho}_s \delta_s}{\overline{\rho}_e \delta_e}.
\ee
Note that, in the cosmological constant case, $\delta_e = 0$. So, when $n=0$, $c_{s,T}^2 = c_{s,s}^2 = w_s$. We also find that in the post-lingering phase of evolution, $\overline{\rho}_s a^n \ll \overline{\rho}_e a^m$, so $Y \rightarrow 0$, and $c_{s,T}^2 \rightarrow w_e$. In the lingering phase, however, we have that
\be
a^{n-m} \frac{\overline{\rho}_s}{\overline{\rho}_e} = \frac{2-n}{m-2}\frac{1+\Delta_s}{1+\Delta_e}(1+\Delta)^{n-m} \simeq \frac{2-n}{m-2},
\ee
when $n\neq2$, and
\be
a^{n-m} \frac{\overline{\rho}_s}{\overline{\rho}_e} = \frac{\kappa^2 \Delta\rho_s}{3} \frac{(a_\ast(1+\Delta))^{2-m}}{1+\Delta_e} \simeq -\Delta_e,
\ee
when $n=2$. In the above manipulations, we keep the lowest order terms after using \eqref{Hub1O1} and \eqref{Hub1N2O1}. Thus, we can say that $Y < \mathcal{O}(1) \delta_s/\delta_e$. Thus, the sound speed characterizing the combination of fluids in the universe remains constant only if the ratio of perturbations in the respective fluids remains constant.

\subsection{Background quantities of interest}\label{param}
To determine the metric perturbation in the lingering phase, we recall \eqref{LoitScaleGen}, \eqref{LoitScale2}, and the fact that $\Delta_e$ is a small quantity. The background evolution enters into the perturbation equations as
\begin{equation}
\begin{array}{rcl}
    \mathcal{D}_n^l & = & 1 - \frac{\Delta_e}{2} \big( (m-2) \cosh{\sqrt{\frac{1}{2}(m - 2)(2 - n)} \eta}\big) \\[10pt]
    z_n^l & = & \frac{\sqrt{3(m - 2)(2 - n)}\Delta_e}{2 a_\ast(2 - n)} \sinh{\sqrt{\frac{1}{2}(m - 2)(2 - n)} \eta} \\[10pt]
    \frac{z_n^{l''}}{z_n^l} & = & \frac{1}{2}(m - 2)(2 - n) - \frac{3}{8} \Delta_e \big( (m - 2) ( m (n - 2) - 2 (n - 6) ) \cosh{\sqrt{\frac{1}{2}(m - 2)(2 - n)} \eta} \big) \\[10pt]
    \mathcal{D}_2^l & = & 1 - \frac{1}{2}(m - 2) \Delta_e \\[10pt]
    z_2^l & = & \sqrt{\frac{3}{2}} \frac{(m - 2) \eta}{2 a_\ast} \Delta_e \\[10pt]
    \frac{z_2^{l''}}{z_2^l} & = & - 3 (m - 2) \Delta_e
\end{array}
\end{equation}
where the subscript $l$ stands for lingering. When the exotic fluid dominates, we perform the same calculations. Denoting the pre-factor of the scale factor in \eqref{PostLoitScaleGen} $d$, we find that
\begin{align}
   \mathcal{D}_n^{pl} &= \frac{n}{2 \sin(\frac{n-2}{2}\Delta\eta_{pl} + \eta_1)^2}, \\
   z_n^{pl} &= \frac{\sqrt{3}}{d \sqrt{n}} \cos(\frac{n-2}{2}\Delta\eta_{pl} + \eta_1) \sin(\frac{n-2}{2}\Delta\eta_{pl} + \eta_1)^{2/(2-n)}, \\
   \frac{z^{{pl}''}_n}{z_n^{pl}} &= \frac{1}{4} \Bigg(\frac{2n}{\sin(\frac{n-2}{2}\Delta\eta_{pl} + \eta_1)^2} - (n - 4)^2 \Bigg) \\
   \mathcal{D}_2^{pl} =& 1 + \Delta_e, \\
    z_2^{pl} =& - a_2^{-1} \sqrt{\frac{3}{2}(1 - \Delta_e) \Delta_e}, \\
    \frac{z_2^{{pl}''}}{z_2^{pl}} =& \Delta_e.
\end{align}
\subsection{The case of $n=2$}
It turns out that $n=2$ actually simplifies our perturbation calculations: $z''/z$ is constant in both phases of scale factor evolution in that case. The solution to the perturbation equation for both phases is then 
\be
u = C_1 e^{i A_{(pl)l} \Delta\eta} + C_2 e^{-i A_{(pl)l} \Delta\eta},
\ee
where
\begin{align}
A_l^2 &= c_{s,T}^2 (k^2 - 1) + 3 (m - 2) \Delta_e, \\
A_{pl}^2 &= w_e (k^2 - 1) - \Delta_e.
\end{align}
The subscript $l$ denotes lingering and $pl$ post-lingering. For $n=2$, $w_e = -1/3$. We see that the solutions become unstable for all $k$ in the post-lingering phase. 

During the lingering phase, the stability depends on the sound speed. For $Y \gg 1$, the standard matter perturbations dominate and the equation of state (and so also sound speed) approach $w_s \geq 0$. Since $\Delta_e$ is assumed positive and $m \geq 3$, the metric perturbation oscillates for all $k$ in this limit. When $Y \ll 1$, the sound speed squared is negative. Thus, the frequency of oscillations becomes complex for
\be
k^2 - 1 > 9 (m - 2) \Delta_e.
\ee
Since $k \in \mathbb{N}$, $k^2 -1 \geq 0$. If $\Delta_e > 0$, there will always exist at least one mode for which the above inequality is violated: perturbations on the 3-curvature scale remain stable and oscillatory. Modes of any smaller scale will exponentially increase or decrease.

The metric perturbation $\Phi$ is obtained from the above solutions by using the appropriate $\mathcal{D}$'s. 

\begin{figure}
    \centering
    \begin{subfigure}[b]{.5\textwidth}
         \centering
\includegraphics[width=\textwidth]{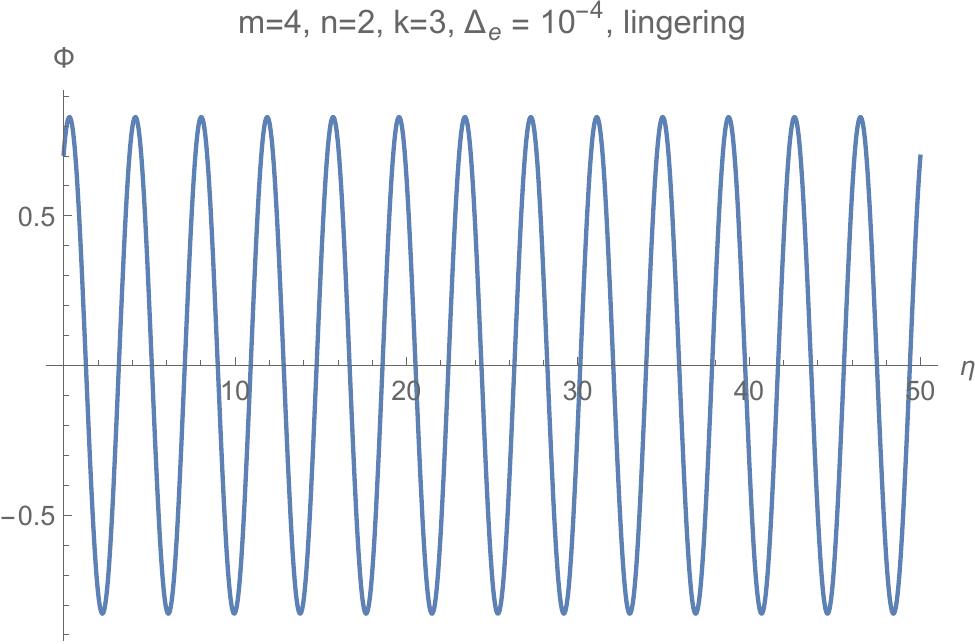}
    \end{subfigure}%
    \begin{subfigure}[b]{.5\textwidth}
         \centering \includegraphics[width=\textwidth]{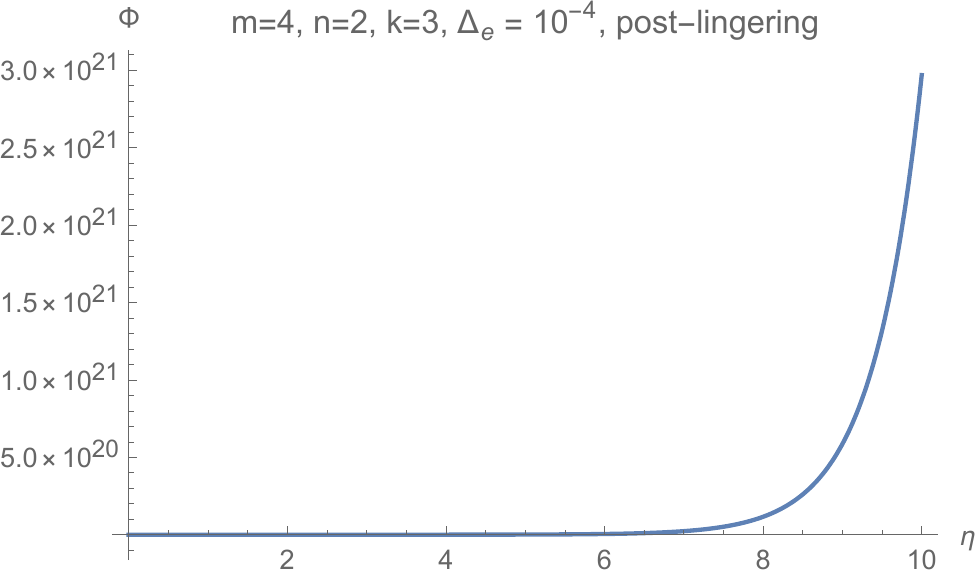}
    \end{subfigure}
    \caption{ Evolution of $\Phi$ (measured in units of $\kappa$) for a universe composed of radiation and curvature.}
    \label{n2r}
\end{figure}

Let us consider the case for $m=4$ (radiation). One finds that for the potential $\Phi$ oscillates during the lingering phase, and grows during the post-lingering phase. Fig. \ref{n2r} shows the time evolution for $k=3$. The lingering period lasts for a conformal time interval $\Delta\eta\sim140$. The scale factor grows exponentially during the post-lingering period. 

\subsection{Lingering}
\begin{figure}
    \centering
    \includegraphics[width=.65\textwidth]{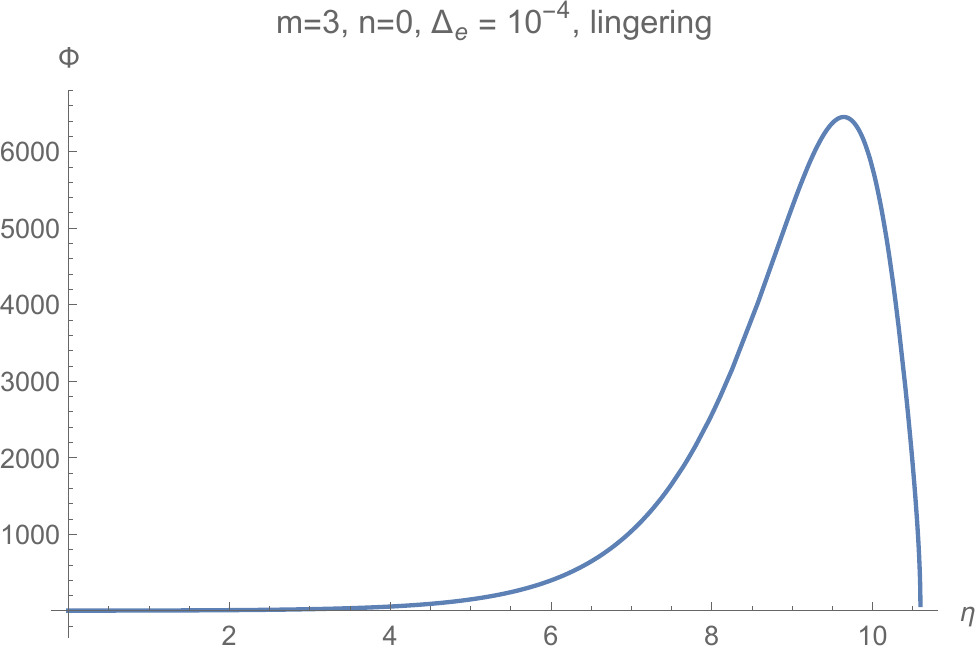}
    \caption{Evolution of $\Phi$ (measured in units of $\kappa$) for a universe composed of matter and cosmological constant.}
    \label{matccpert}
\end{figure}

\begin{figure}
    \centering
    \includegraphics[width=.65\textwidth]{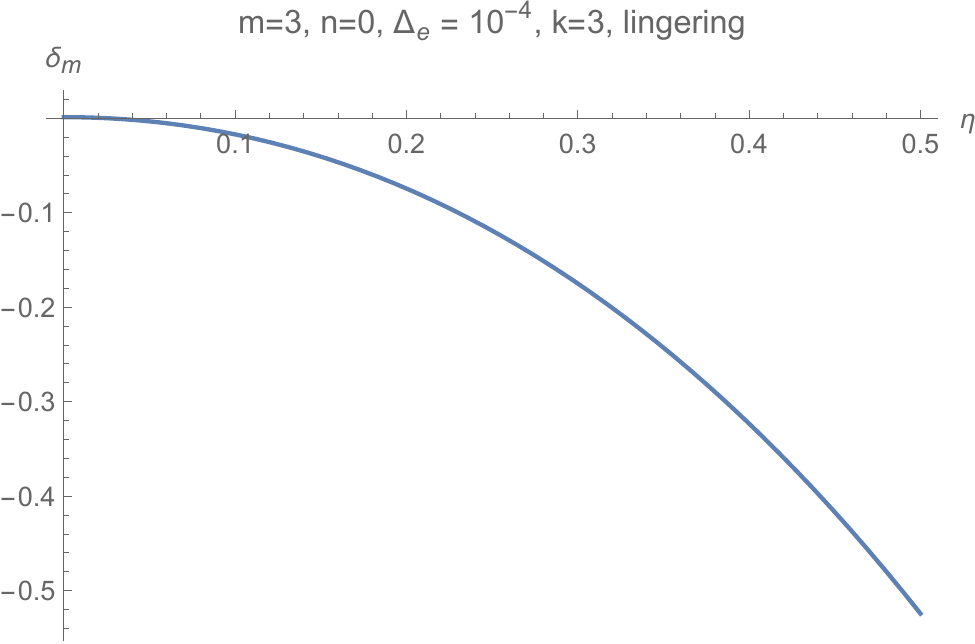}
    \caption{Evolution of $\delta_m$ during lingering for a universe composed of matter and cosmological constant.}
    \label{matdenl}
\end{figure}

Solutions to the perturbation equation during the lingering phase for general $n$ is given by
\begin{align}
   \text{u}_n^l (\eta) &= C_1 \text{MathieuC}\left[\frac{8 \text{c}_s^2 \left(k^2-1\right)}{(m-2) (n-2)}+4,\frac{3 \text{$\Delta $}_e (n-6)}{n-2}-\frac{3 \text{$\Delta $}_e m}{2},\frac{1}{2} i \eta \sqrt{-\frac{m n}{2}+m+n-2}\right]\\
   &-C_2 \text{MathieuS}\left[\frac{8 \text{c}_s^2 \left(k^2-1\right)}{(m-2) (n-2)}+4,\frac{3 \text{$\Delta $}_e (n-6)}{n-2}-\frac{3 \text{$\Delta $}_e m}{2},\frac{1}{2} i \eta \sqrt{-\frac{m n}{2}+m+n-2}\right] ,
\end{align}
where $C_1$ and $C_2$ are constants to be fixed by the initial conditions. 
To get a feel for the solutions, we solve the perturbation equation numerically for certain combinations of fluids, assuming $u(0) = 1$ and $u'(0) = 0$.\\
\\
For matter and cosmological constant, $c_s^2 = 0$, since $w_s = 0$ and $\delta_e = 0$. This implies the perturbation has no $k$ dependence. Figure \ref{matccpert} is a plot of the solution, the lingering phase lasts over a conformal time interval $\Delta\eta \approx 15$. $\Phi$ can be obtained from $u$ by using $\mathcal{D}^l_0$. The curvature perturbation $\mathcal{R} \approx -\Phi$ during lingering.
\\
\\
The evolution of the density contrast is given by Eq. \ref{denconsttime} specialized to the case of a universe composed of matter and cosmological constant
\begin{align}
\delta^{''}_{m}+\Hcon\delta_{m}^{'} +(k^2-1)\Phi-3\Hcon\Phi^{'}-3\Phi^{''}&=0,
\end{align}
there is no perturbation in a cc-like fluid. Fig. \ref{matdenl} shows the time evolution of the density contrast for $k=3$ assuming $\delta_{r}(0)=\delta_{r}^{'}(0)=0.001$

\begin{figure}
    \centering
    \includegraphics[width=.65\textwidth]{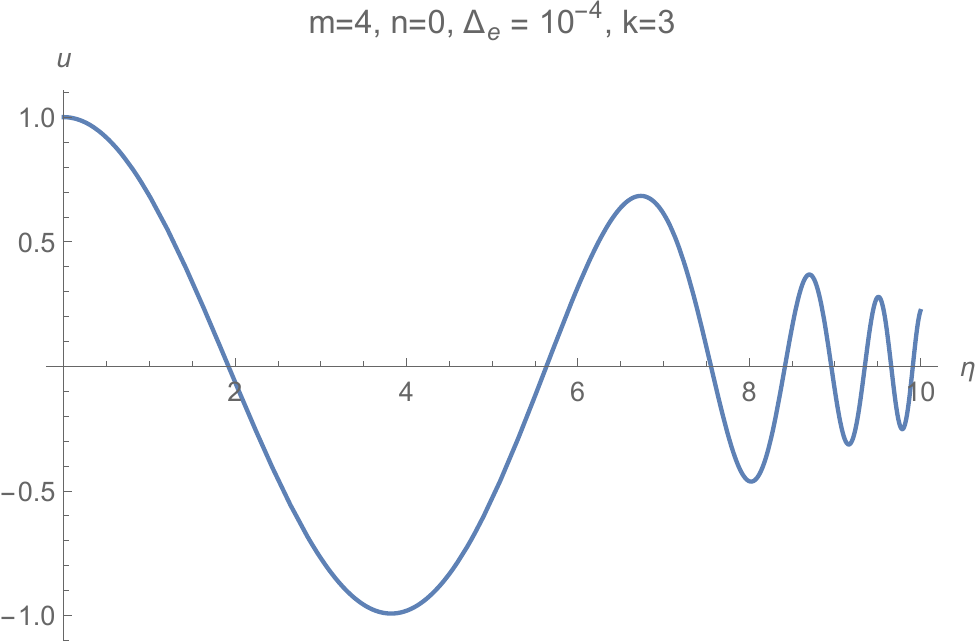}
    \caption{Evolution of $u$ during lingering for a universe composed of radiation and cosmological constant.  }
    \label{radccpert}
\end{figure}

\begin{figure}
    \centering
    \includegraphics[width=.65\textwidth]{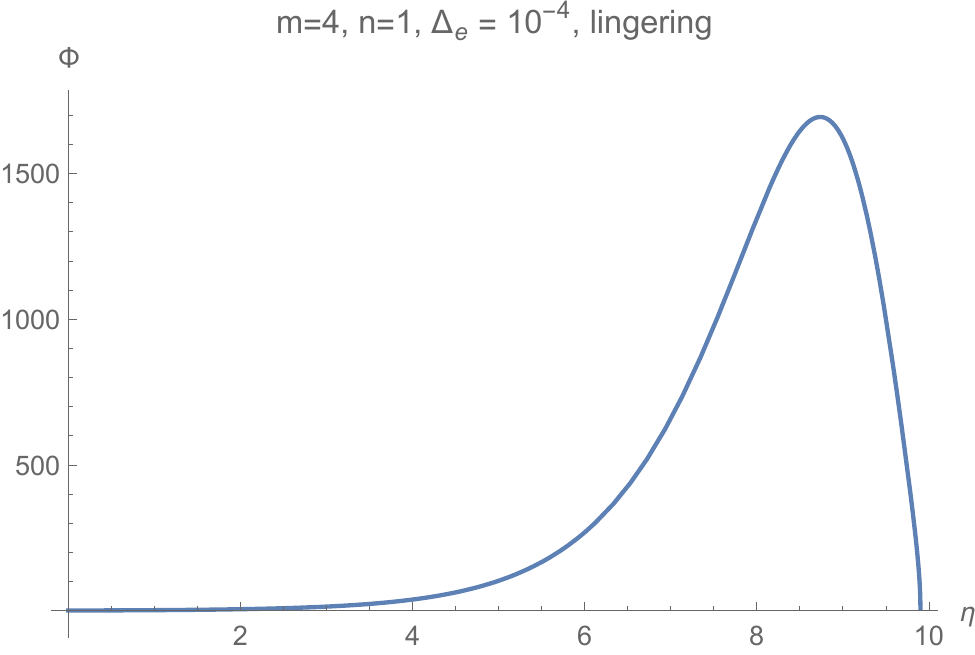}
    \caption{Evolution of $\Phi$ (in units of $\kappa$) for a universe composed of radiation and string network.}
    \label{rsnl}
\end{figure}

\begin{figure}
    \centering
    \begin{subfigure}[b]{.5\textwidth}
         \centering
\includegraphics[width=\textwidth]{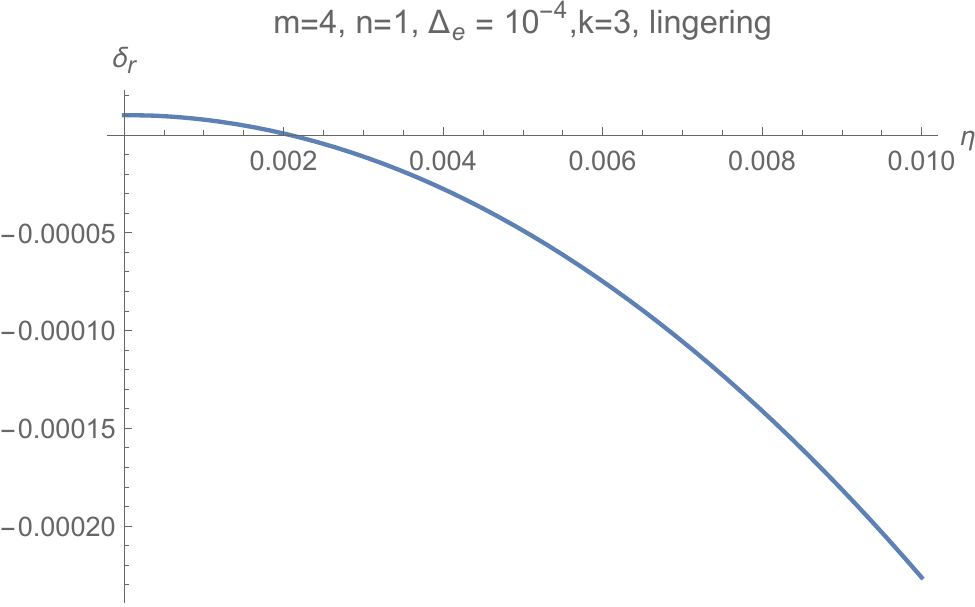}
    \end{subfigure}%
    \begin{subfigure}[b]{.5\textwidth}
         \centering \includegraphics[width=\textwidth]{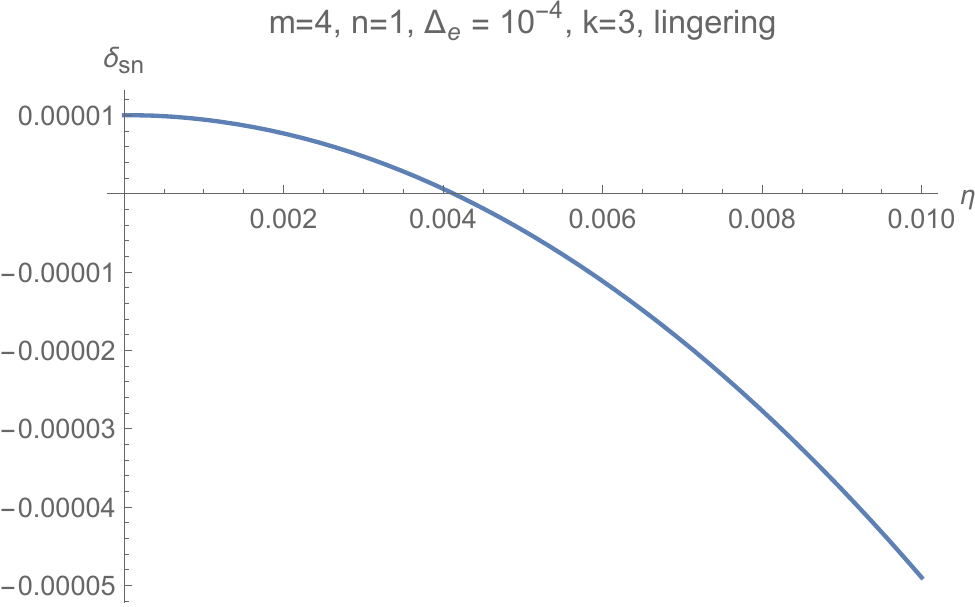}
    \end{subfigure}
    \caption{Evolution of density contrasts ($\delta_{r,sn}$) for a universe composed of radiation and string network. The growth/decay is quick in conformal time.}
    \label{rsndenl}
\end{figure}

The Hubble radius diverges in the lingering phase, so all modes start out sub-Hubble. For small $\Delta_e$ and $\eta \rightarrow 0$ the perturbation equation for matter plus cosmological constant can be approximated by
\be
u'' - u = 0
\ee
The initial quantum conditions would correspond to quantizing an inverted harmonic oscillator with potential $V = -\frac{1}{2}u^2$. \\
\\
Let us now look at universe composed of radiation and a cosmological constant-like fluid, $c_s^2 = 1/3$, since $w_s = 1/3$ and $\delta_e = 0$. The perturbations now have $k$ dependence. Figure \ref{radccpert} is a plot of the solution, the lingering phase lasts over a conformal time interval $\Delta\eta \approx 10$.
\\
\\
For small $\Delta_e$ and $\eta \rightarrow 0$ the perturbation equation can be approximated by
\be
u'' + \Big(\frac{1}{3}(k^2-1) - 2\Big)u = 0
\ee
This gives oscillatory solutions for $k\geq 3$.\\
\\
Let us do a similar analysis for radiation and string network. Assuming adiabatic initial conditions for the perturbations we get $\delta_r=4\delta_{sn}$ and the effective sound speed then depends only on the relative energy density between the two sectors which is approximately constant during lingering. Interestingly, we find $c_s^2 = 0$ during lingering for radiation and string network, which implies there is no $k$ dependence. The lingering phase lasts over a conformal time interval $\Delta\eta \sim 10$. Using $\mathcal{D}_{1}$ and $\frac{z_1^{l''}}{z_1^l}$ from section \ref{param}, we numerically solve eqs. \ref{phiconv} and \ref{ptbeq} with the initial condition $u_1(0) = u'_1(0) =1$. Fig. \ref{rsnl} is a plot of the conformal time evolution of $\Phi$.
\\
\\
The evolution of the density contrasts is given by Eq. \ref{denconsttime} specialized to the case of radiation and string network
\begin{align}
    \delta^{''}_r +(k^2-1)\frac{\delta_{r}}{3}+\frac{4}{3}(k^2-1)\Phi-4\Phi^{''}&=0\\
\delta^{''}_{sn}+3\Hcon\delta_{sn}^{'} -(k^2-1)\frac{2\delta_{sn}}{3}+\frac{(k^2-1)}{3}\Phi-3\Hcon\Phi^{'}-\Phi^{''}&=0.
\end{align}
It is interesting to note that the evolution of $\delta_r$ has no explicit $\Hcon$ dependence and the choice of positive curvature results in growing/decaying solutions. Fig. \ref{rsndenl} shows the time evolution of the density contrasts for $k=3$, assuming $\delta_{r,sn}(0)=\delta_{r,sn}^{'}(0)=10^{-5}$.
\\
\\
We now connect the perturbation at the end of the lingering phase to the post lingering solution.

\subsection{Post-lingering}
Let's now turn to the analysis of the metric perturbation in the post-lingering phase for general $n$. Via examination of $z_n''/z_n$ for this phase, we find that it's reasonable to take the argument of the sine function to be close to $\pi$. This is where the scale factor diverges and it is during this period of expansion that most momentum modes exit the Hubble radius. We can then write
\be
\frac{z_n''}{z_n} \simeq \frac{2n}{(n-2)^2(\eta-c)^2} - \frac{1}{12}(n-6)(3n-8).
\ee
This means the metric perturbation equation becomes
\be
u_n'' + (A - \frac{B}{(\eta-c)^2})u_n =0,
\ee
where
\be
A = c_{s,T}^2 (k^2 - 1) + \frac{1}{12}(n-6)(3n-8), \quad B = \frac{2n}{(n-2)^2}, \quad c = \frac{2(\pi - \eta_1)}{n-2}.
\ee
We can write solutions thereof in terms of Whittaker functions:
\be
u = C_1 M_{0,\mu}(x) + C_2 W_{0,\mu}(x),
\ee
where we defined $2\mu = \sqrt{1+4B}$ and $x = 2 i \sqrt{A} (c-\eta)$. We now get to go on a fun math digression. Whittaker functions can be written in terms of the confluent hypergeometric functions, $M(a,b;x)$ and $U(a,b;x)$:
\be
M_{k,\mu}(x) = e^{-x/2} x^{\mu+1/2} M( \mu - k + 1/2, 2\mu + 1; x),
\ee
and the corresponding equation for the second Whittaker function can be found by taking $M_{k,\mu} \rightarrow W_{k,\mu}$ and $M(a,b;x) \rightarrow U(a,b;x)$. Most convenient for our situation, the hypergeometric functions can be replaced by elementary functions for integer scaling of the energy densities. Using the full expression for $\mu$, we find that
\be
u_n = e^{-x/2} x^{2/(2-n)} \bigg[C_1 M\bigg(\frac{2}{2-n},\frac{4}{2-n};x\bigg) + C_2 U\bigg(\frac{2}{2-n},\frac{4}{2-n};x\bigg) \bigg].
\ee
Our general-n solutions apply to $n=0$ and $n=1$. Thus, we can specify that
\begin{align}
    M(1,2;x) &= \frac{e^x-1}{x} &\quad U(1,2;x) &= \frac{1}{x} \\
    M(2,4;x) &= \frac{6(2 + x + e^x (x - 2))}{x^3} &\quad U(2,4;x) &= \frac{2 + x}{x^3}. 
\end{align}

Now let's transform back to the metric perturbation. As covered above, there are problems with $\mathcal{D}_n$ when $n=0$:
\be
\frac{\sqrt{\mathcal{D}_n}}{a_n} = \pm\frac{1}{a_{pl}} \sqrt{\frac{n}{2}} \sin\bigg(\frac{n-2}{2} \eta + \eta_1 \bigg)^{\frac{2}{2-n}}\sin( \eta_1)^{\frac{2}{n-2}} .
\ee
Clearly, this goes to zero for an exact cosmological constant, which invalidates the transformation to $\Phi$. However, we have to remember that we solve the perturbation equation during a specific regime of cosmic evolution. Taking the argument of sine to be close to $\pi$, we get
\be
\frac{\sqrt{\mathcal{D}_n}}{a_n} \simeq \frac{\sqrt{n}}{a_{pl}}\sin( \eta_1)^{\frac{2}{n-2}}(2-n)\bigg(\frac{-x}{2i\sqrt{A}}\bigg)^{\frac{n}{2-n}}\equiv d x^{\frac{n}{2-n}} 
\ee
Therefore, our metric perturbation follows
\be
\Phi_n = \frac{\kappa}{\sqrt{2}} d e^{-x/2} x^{\frac{2+n}{2-n}} \bigg[C_1 M\bigg(\frac{2}{2-n},\frac{4}{2-n};x\bigg) + C_2 U\bigg(\frac{2}{2-n},\frac{4}{2-n};x\bigg) \bigg].
\ee
Note that, in the solution above, every term in the product (except the very first) has $n$-dependence in it. We will have to do a case--by-case analysis for the different fluids to learn more about these solutions.

\begin{figure}
    \centering
\includegraphics[width=.65\textwidth]{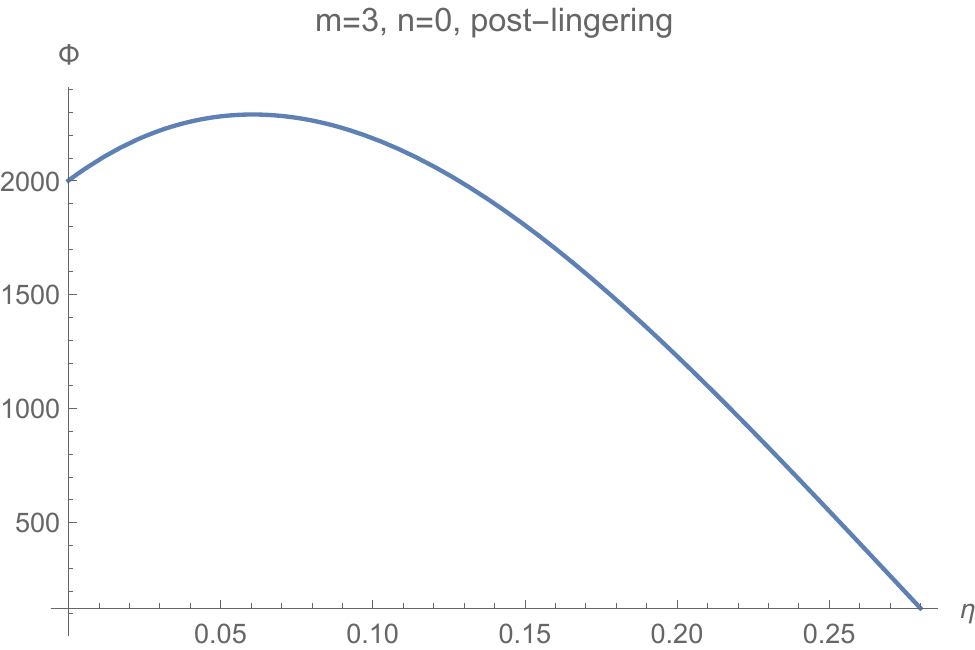}
    \caption{Post-lingering evolution of $\Phi$ (in units of $\kappa$) for a universe composed of matter and a cosmological constant-like fluid.}
    \label{matccpl}
\end{figure}
\begin{figure}
    \centering
\includegraphics[width=.65\textwidth]{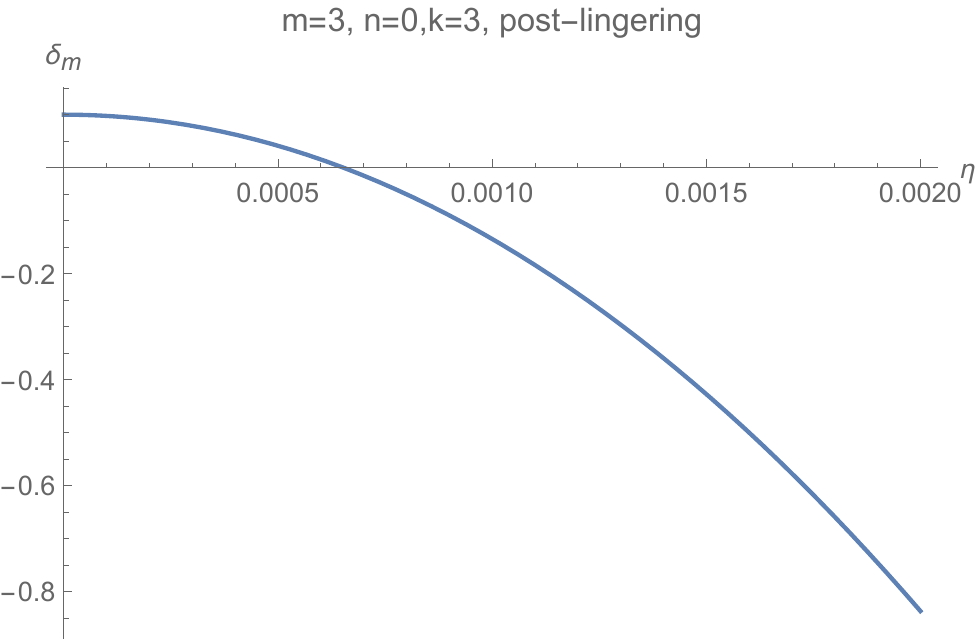}
    \caption{Evolution of $\delta_m$ for a universe composed of matter and cosmological constant.}
    \label{matccdenpl}
\end{figure}
\begin{figure}
    \centering
\includegraphics[width=.65\textwidth]{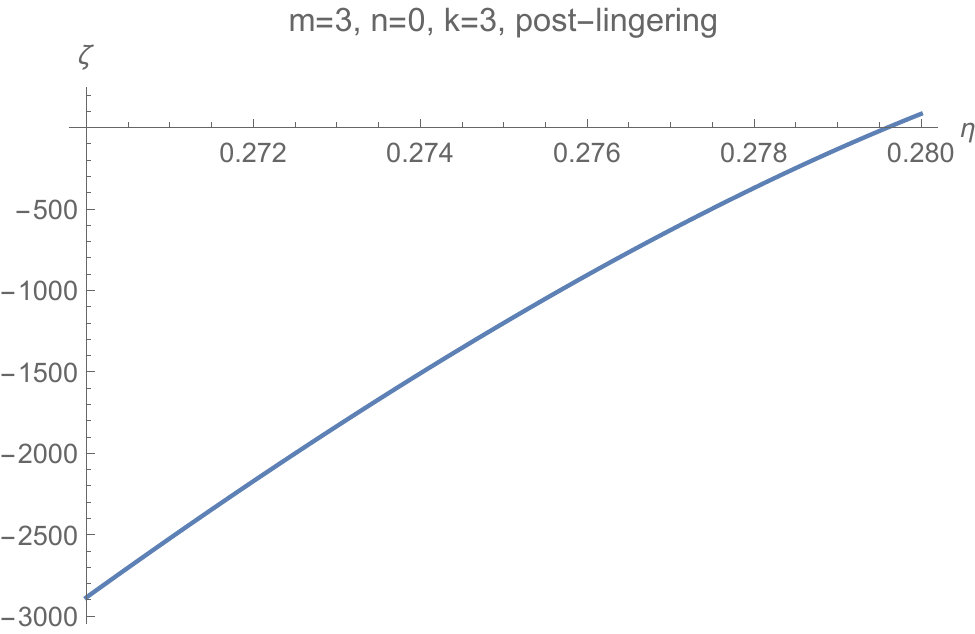}
    \caption{Evolution of $\zeta$ for a universe composed of matter and cosmological constant.  }
    \label{zetamatccpl}
\end{figure}
\begin{figure}
    \centering
\includegraphics[width=.65\textwidth]{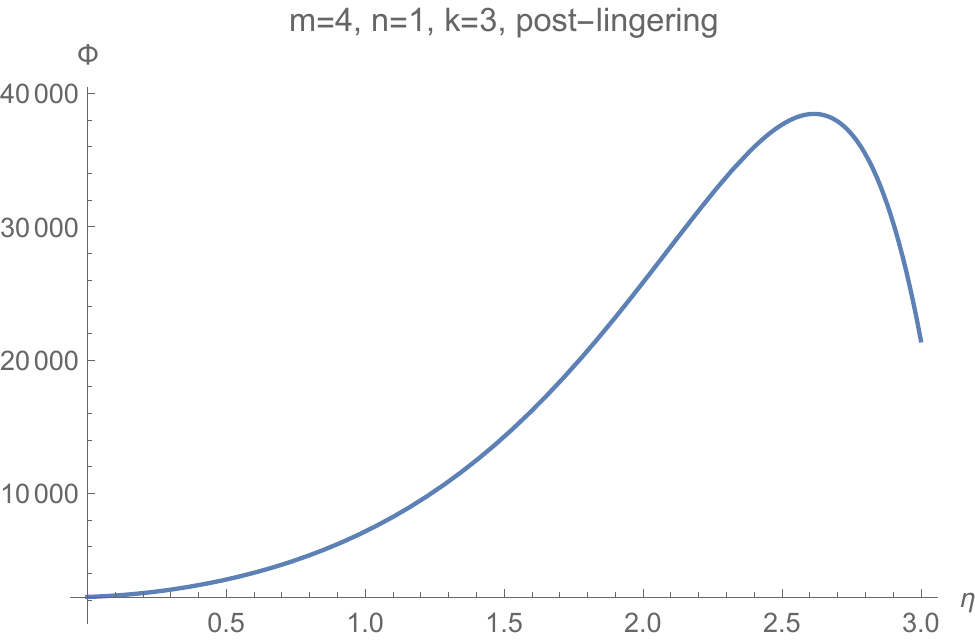}
    \caption{Evolution of $\Phi$ (in units of $\kappa$) for a universe composed of radiation and string network.}
    \label{rsnpl}
\end{figure}
\begin{figure}
    \centering
    \begin{subfigure}[b]{.5\textwidth}
         \centering
\includegraphics[width=\textwidth]{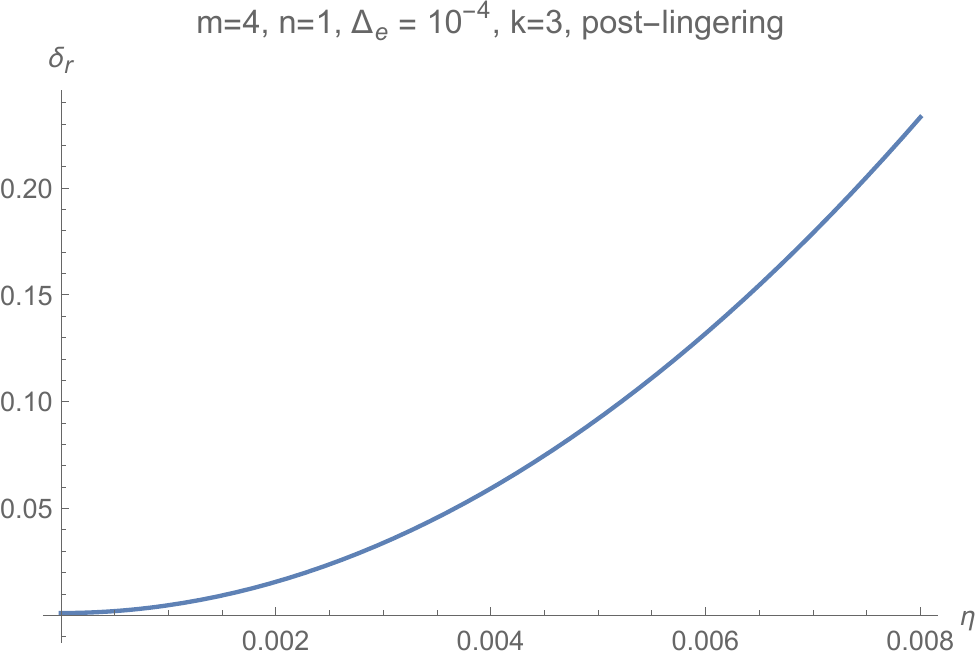}
    \end{subfigure}%
    \begin{subfigure}[b]{.5\textwidth}
         \centering \includegraphics[width=\textwidth]{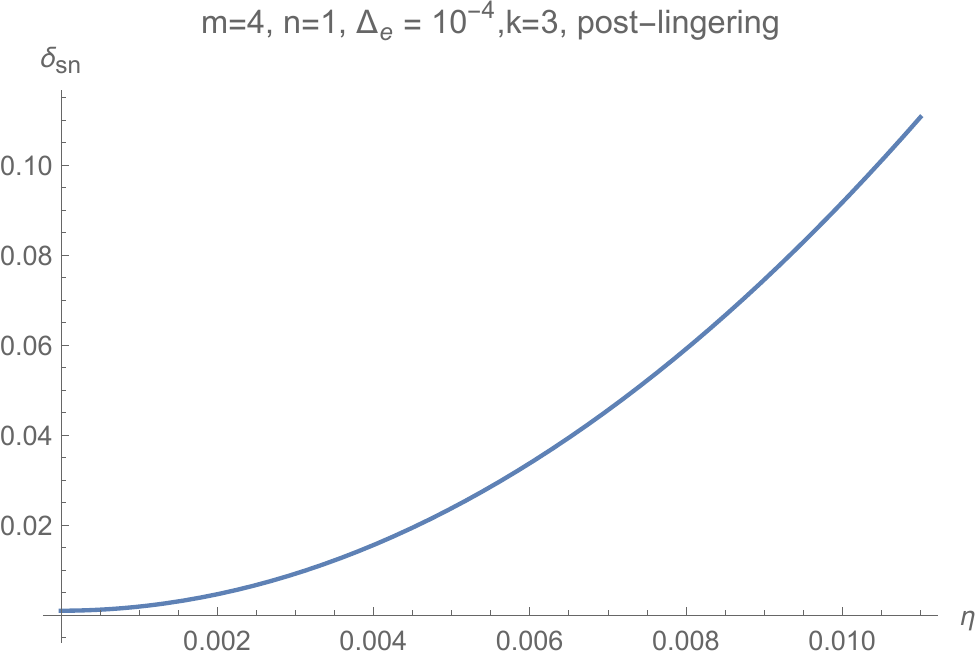}
    \end{subfigure}
    \caption{ Evolution of $\delta_{r,sn}$ in a universe composed of radiation and string-network. }
    \label{rsndenpl}
\end{figure}
\begin{figure}
    \centering
\includegraphics[width=.65\textwidth]{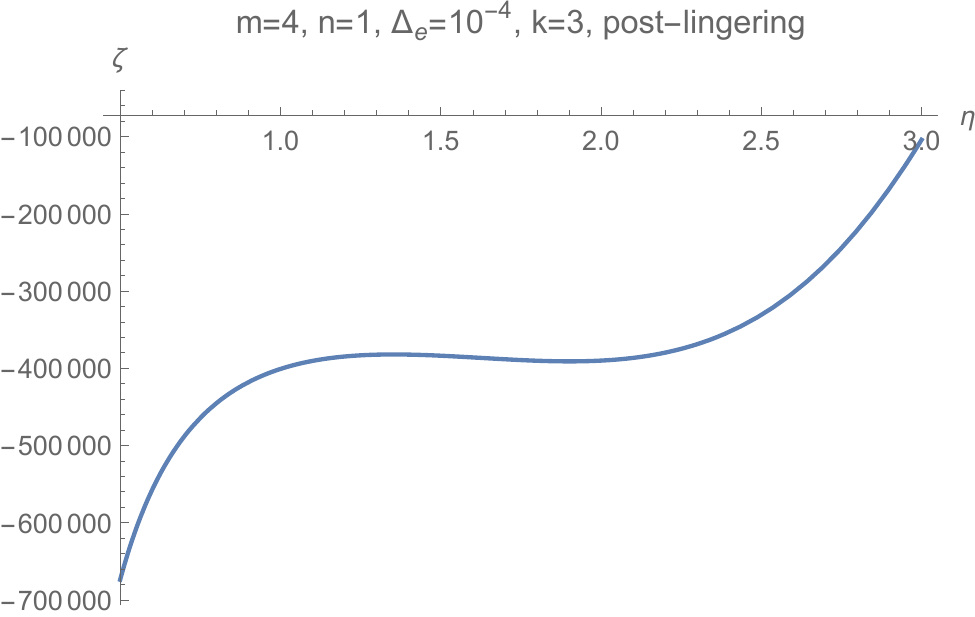}
    \caption{Evolution of $\zeta$ in a universe composed of radiation and string-network.}
    \label{zetarsnpl}
\end{figure}

Let us analyze the perturbations numerically for matter and a cosmological constant-like fluid with certain simplifying approximations to get a feel for the evolution. For a cosmological constant dominated universe $\rho \sim \text{const.} \sim \frac{1}{\kappa^2} \text{(lingering value)}$. Deep into the post-lingering phase the scale factor diverges, hence the curvature term in the first Friedmann equation (7) can be ignored. We then find the approximate solution
 \begin{equation}
     a = \frac{2}{1-2\sqrt{3}\eta} \implies \mathcal{H} = \frac{2\sqrt{3}}{1-2\sqrt{3}\eta}
 \end{equation}
where we use $a(0)\sim2$ as the start of the post lingering phase, we also note that $a$ and $\mathcal{H}$ diverges around $\eta\sim \frac{1}{2\sqrt{3}}$. Equation \ref{GenEinCombo} for cosmological constant domination simplifies to
\be
\label{ccptb}
\Phi'' + 3 \Hcon \Phi' + \left(3 \Hcon^2  +5\right) \Phi = 0, 
\ee
where we used $c_{s,T}^2=0$ and
\be
w_{eff} = -\Omega_e \sim -(1+\Omega_K) = -\bigg(1+\frac{1}{\Hcon^2}\bigg).
\ee
Eq. \ref{ccptb} corresponds to an underdamped harmonic oscillator. Figure \ref{matccpl} shows the evolution of $\Phi$ assuming $\Phi(0)=2000$ and $\Phi'(0)=10^4$.
\\
\\
Fig. \ref{matccdenpl} shows the time evolution of the density contrasts for $k=3$, assuming $\delta_{m}=\delta_{m}^{'}=0.1$ at the transition from lingering to post lingering.
The evolution of $\zeta$ is shown in Fig.\ref{zetamatccpl}.
\\
\\
We do a similar numerical analysis for radiation+string network. $c_s^2 \sim w_{sn}=-2/3$. The perturbation evolves with time according to
\be
\label{rsnpleq}
u_1'' + \left(-\frac{2}{3}(k^2 - 1) - \bigg(\frac{2}{\eta +\pi}-\frac{25}{12}\bigg)\right)u_1 = 0,
\ee
where we took $\eta_1\sim \pi/2$. The potential $\Phi$ is obtained from $u_1$ by the multiplicative factor
\be
\frac{\sqrt{D_1}}{a_1} \sim \frac{1}{2\sqrt{2}}\bigg(\frac{\pi-\eta}{2}\bigg).
\ee
Fig. \ref{rsnpl} is a plot of the time evolution of the potential $\Phi$ for $k=3$. The scale factor diverges around $\eta \sim \pi$. We choose $u_1(0)=4000$ and $u'_1(0)=2500$ to connect with the lingering phase.
\\
\\
Fig. \ref{rsndenpl} shows the time evolution of the density contrasts for $k=3$, assuming $\delta_{r,sn}=\delta_{r,sn}^{'}=0.001$ at the transition from lingering to post lingering.
The evolution of $\zeta$ is shown in Fig. \ref{zetarsnpl}.

\section{Conclusions}

 We again emphasize that there is some disagreement in the initial singularity problem of inflation as first established by BGV in \cite{Borde:2001nh} by showing the past geodesic incompleteness of inflation. And this theorem has been criticized in several recent works \cite{Geshnizjani:2023hyd,Easson:2024uxe}, whereas other authors argue differently \cite{Kinney:2023urn}. This is an interesting issue to investigate further, and although this was part of our motivation, we also think it is worthwhile to pursue an alternative beginning to inflation where again we can avoid troubles of quantum field theory in deSitter space (e.g. defining the S-matrix) and could also lead to transplanckian physics. Most importantly, our primary motivation is that this provides an alternative to a sudden inflationary period (or eternal inflation) and that of an always evolving (or cyclic) universe. Instead the universe began in a static state that as we show here was unstable to inflation.  Philosophically, this provides a third paradigm for how the universe began. Much work remains...

\section*{Acknowledgements}
We thank  Amy Burks, Damien Easson, Ghazal Geshnizjani and Will Kinney for useful conversations. We especially thank Kenny Ratliff and Kuver Sinha for initial collaboration. S.W. thanks KITP Santa Barbara, Alexey Petrov and the University of South Carolina and the Simons Center for hospitality. 
This research was supported in part by DOE grant DE-FG02-85ER40237.

\bibliographystyle{apsrev4-1}
\bibliography{paper.bib}

\end{document}